\documentclass[]{spie}  

\usepackage{amsmath,amsfonts,amssymb,amsopn}
\usepackage{graphicx}
\usepackage{xcolor}
\usepackage{cite}
\usepackage{xspace}
\usepackage{subfig} 
\usepackage{mathtools} 
\usepackage{url}

\newcommand{\bv}[1]{\boldsymbol{#1}}
\newcommand{\dfn}{\coloneqq}
\newcommand{\untsph}{\mathbb{S}^{2}} 

\newcommand{\shc}[3]{({#1})_{#2}^{#3}}
\newcommand{\lsph}{L^2(\untsph)}
\newcommand{\conj}[1]{\overline{#1}} 
\newcommand{\unit}[1]{\bv{\hat{#1}}}

\newcommand{\intsph}{\int_{\untsph}}

\newcommand{\figref}[1]{Fig.\,\ref{#1}}

\newcommand{\appref}[1]{Appendix\,\ref{#1}}
\newcommand{\secref}[1]{Section\,\ref{#1}}

\newcommand{\rotopz}{\mathcal{D}(\unit{y})}
\newcommand{\snr}{\textrm{SNR}}
\newcommand{\qed}{\hfill \ensuremath{\Box}}

\DeclarePairedDelimiterX\abs[1]{\lvert}{\rvert}{#1}
\DeclarePairedDelimiterX\parn[1]{(}{)}{#1}
\DeclarePairedDelimiterX\set[1]{\lbrace}{\rbrace}{#1}
\DeclarePairedDelimiterX\innerp[2]{\langle}{\rangle}{#1,#2}
\DeclarePairedDelimiterX\norm[1]{\lVert}{\rVert}{#1}
\DeclarePairedDelimiterX\brak[1]{\lbrace}{\rbrace}{#1}

\DeclarePairedDelimiterX\explim[1]{\lbrace}{\rbrace}{#1} 
\DeclareMathOperator{\expectop}{\mathbb{E}} 
\newcommand{\expect}[2][]{\expectop\explim[#1]{#2}}

\graphicspath{{figs/},{pdf/},{Figures/}}

\newtheorem{remark}{Remark}

\title{Spatio-spectral Formulation and Design of Spatially-Varying Filters for Signal Estimation on the 2-Sphere}

\author{Zubair Khalid, Rodney A. Kennedy, Parastoo Sadeghi and
Salman Durrani \skiplinehalf  Research School of Engineering, The
Australian National University, Canberra, Australia}

\authorinfo{Email: \{zubair.khalid,~rodney.kennedy,~parastoo.sadeghi,~salman.durrani\}@anu.edu.au}

\begin{document}

\maketitle

\begin{abstract}

In this paper, we present an optimal filter for the enhancement or
estimation of signals on the 2-sphere corrupted by noise, when
both the signal and noise are realizations of anisotropic processes on
the 2-sphere. The estimation of such a signal in the spatial or
spectral domain separately can be shown to be inadequate. Therefore, we develop an optimal
filter in the joint spatio-spectral domain by using a framework recently
presented in the literature --- the spatially localized spherical harmonic
transform --- enabling such processing.  Filtering of a signal in the
spatio-spectral domain facilitates taking into account anisotropic
properties of both the signal and noise processes. The proposed
spatio-spectral filtering is optimal under the mean-square
error criterion. The capability of the proposed filtering framework
is demonstrated with by an example to estimate
a signal corrupted by an anisotropic noise process.
\end{abstract}



\section{Introduction}

The development of signal processing techniques for signals defined
on the 2-sphere finds many applications in various fields of
science and engineering. These applications include surface analysis
in medical imaging~\cite{Chung:2010}, geodesy and planetary
studies~\cite{Simons:1997,Wieczorek:2005,Audet:2011}, the study and
analysis of cosmic microwave background~(CMB) in
cosmology~\cite{McEwen:2007:cosmology,Wiaux:2005}, 3D
beamforming~\cite{Ward:1995} and wireless channel
modeling~\cite{Pollock:2003} in communication systems. In this work,
we consider the problem to design optimal filters for the
enhancement or estimation of signals, defined on the 2-sphere,
corrupted by an \emph{anisotropic} noise process.

The problem to design optimal filters for signals defined on the
sphere has been well studied\cite{Klees:2008,McEwen:2008,Wei:2011}. At their core, these
investigations assume the signal and/or noise process to be
isotropic and present the formulation of optimal filters in either
spatial~(pixel) domain or spectral domain, which is enabled through the
spherical harmonic transform~\cite{McEwen:2011,kennedy:book}.
Assuming the process to be isotropic, the classical approach to
remove the effect of noise is using the the mean-square error linear
optimal filter~(Wiener filter) formulated in the spectral
domain~\cite{McEwen:2008,Arora:2010}. Since such an optimal filter
makes profit based on the knowledge of the energy distribution of
the noise in the spectral domain and does not take into account the
anisotropic properties of the process, it is not suitable for
filtering out anisotropic noise~\cite{Klees:2008}. The extension of the
mean-square error optimal filter for the anisotropic process results
in a spatially-varying anisotropic filter~(also referred to as an anisotropic
non-symmetric filter~\cite{Klees:2008}) for which a simple
spectral domain formulation cannot be obtained. This motivates us to
design optimal filter in the spatio-spectral domain using the
filtering framework~\cite{Khalid2:2012}, which enables the spatially
varying spectral filtering of signals defined on the 2-sphere. Cognate ideas
exist for the time-frequency filtering of
non-stationary signals using time-varying optimal
filters~\cite{Kirchauer:1995_ICASSP,Mark:1970,Matz:1998}.

\begin{figure*}[t]
    \centering
    \includegraphics[scale=0.85]{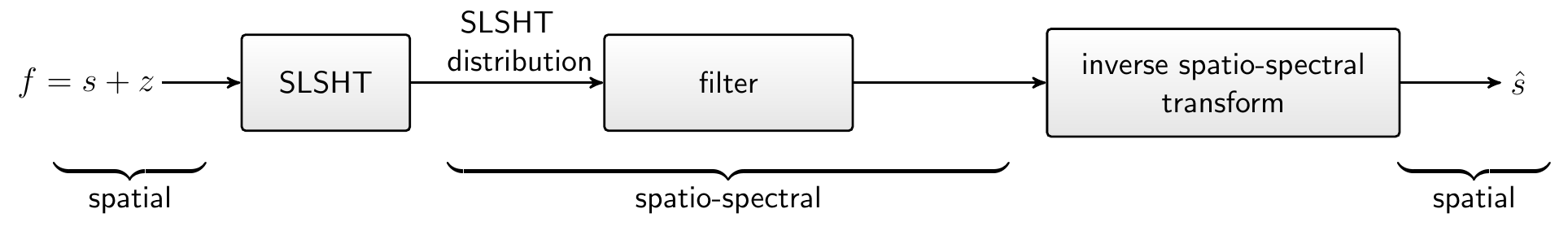}
    \caption{General schematic for spatio-spectral filtering. The spatial
signal, $f = s + z$, on the 2-sphere, is first mapped to the
spatio-spectral domain using the spatially localized spherical
harmonic transform (SLSHT), then its SLSHT distribution  is
transformed in spatio-spectral domain, and the result is mapped back
to the spatial domain using an inverse spatio-spectral transform to
obtain the estimate $\tilde{s}$.}
    \label{fig:block_concept}
\end{figure*}

In this paper, we design optimal filters in the
spatio-spectral domain for the enhancement or estimation of signals
corrupted by the \emph{anisotropic} noise. By employing the
spatially-varying or spatio-spectral filtering methods~\cite{Khalid2:2012} based on the
spatially localized spherical harmonic
transform~(SLSHT)~\cite{Khalid:2012}, we consider the filtering
framework presented in \figref{fig:block_concept}, where the input
signal, $f = s + z$, is a sum of the desired signal $s$ and the
noise $z$ and the output signal $\tilde{s}$ is an estimate of the
desired signal. Given $f$, we serve the objective to find the
estimate $\tilde{s}$ of the desired signal $s$ by first
obtaining the spatio-spectral representation, herein referred as
SLSHT distribution, of the input signal $f$. The SLSHT distribution
is then filtered in the spatio-spectral domain and the inverse
spatio-spectral transform is applied to obtain the signal in the
spatial domain.
The spatio-spectral filtering is made optimal by
choosing the filtering of SLSHT distribution in the spatio-spectral
domain which minimizes the mean square error between the desired
signal $s$ and the estimate $\tilde{s}$. The effectiveness of the
theoretical development is demonstrated with the help of
an example, where we apply proposed optimal filter to enhance a
signal corrupted by anisotropic noise process.

The rest of the paper is organized as follows. In Sections \ref{sec:models} and \ref{sec:ss_filtering},
the mathematical preliminaries and the spatio-spectral
filtering framework are presented, respectively. The problem is
formalized in \secref{sec:problem} and the proposed optimal filter is derived
in \secref{sec:optimal_filtering}. In \secref{sec:example}, an example of optimal estimation of a signal in noise, where both are described by anisotropic processes, is given.  Finally,
the concluding remarks and future directions are given in \secref{sec:conclusions}.

\section{Mathematical Background}
\label{sec:models}

In this section, we briefly review some mathematical background for
signals defined on the 2-sphere or unit sphere.

\subsection{Signals on the 2-sphere}

In this work, we consider the square integrable complex functions
defined on the 2-sphere denoted by $\mathbb{S}^2$. Let
$\unit{x}\equiv \unit{x}(\theta,\phi) \dfn
(\sin\theta\cos\phi,\, \sin\theta\sin\phi,\, \cos \theta)' \in
\mathbb{R}^3$ and $\unit{y}\equiv \unit{y}(\vartheta,\varphi) \dfn
(\sin\vartheta\cos\varphi,\, \sin\vartheta\sin\varphi,\, \cos
\vartheta)' \in \untsph$ denote unit vectors, where $'$
denotes the transpose, Each of the unit vectors parameterizes a
point on the 2-sphere, with $\theta,\,\vartheta \in [0, \pi]$
denoting the co-latitude and $\phi,\,\varphi \in [0, 2\pi)$ denoting
the longitude.

The inner product of two functions $f(\unit{x})$ and $h(\unit{x})$
on $\mathbb{S}^2$ is defined as~\cite{kennedy:book}
\begin{equation}
\label{eqn:innprd}
    \innerp[\big]{f}{h} \dfn  \intsph f(\unit{x})\,\conj{h(\unit{x})} \,ds(\unit{x}),
\end{equation}
where $ds(\unit{x}) = \sin \theta\,d \theta\,d \phi$ is the area
element, $\conj{(\cdot)}$ denotes complex conjugate and the
integration is carried out over the whole sphere. With the
inner product in \eqref{eqn:innprd}, the space of square integrable
complex valued functions on the sphere forms a complete Hilbert
space $\lsph$. Also, the inner product in \eqref{eqn:innprd} induces
a norm $\norm{f} \dfn \innerp{f}{f}{}^{1/2}$. We refer to the
functions with finite induced norm as signals on the 2-sphere.

\subsection{Spectral Representation and Spherical Harmonics}

The Hilbert space $\lsph$ is separable and spherical harmonic
functions (or spherical harmonics for short) $Y_{\ell}^m(\unit{x}) =
Y_{\ell}^m(\theta, \phi)$~\cite{kennedy:book,Sakurai:1994} of all
integer degrees $\ell \ge 0$ and integer orders $-\ell\le m \le
\ell$ form the archetype complete orthonormal set of basis
functions. By completeness, any signal $f\in \lsph$ can be expanded
as
\begin{equation}
\label{Eq:f_expansion}
	f(\unit{x})=\sum_{{\ell}=0}^{\infty}\sum_{m=-{\ell}}^{\ell} \shc{f}{\ell}{m}
	Y_{\ell}^m(\unit{x}),\quad \shc{f}{\ell}{m}\dfn\innerp[\big]{f}{Y_{{\ell}}^m} =
		\intsph f(\unit{x})\conj {Y_{\ell}^m(\unit{x})}\,ds(\unit{x}),
\end{equation}
where $\shc{f}{\ell}{m}$ is the spherical harmonic coefficient of
degree $\ell$ and order $m$, which forms part of the spectral domain
representation of a signal and, therefore, is also referred as
\emph{spectral component}.  The mathematical definition for the spherical
harmonics is provided in \appref{App:maths}.  The
signal $f(\unit{x})$ is said to be band-limited with maximum
spectral degree $L_f$ if $\shc{f}{\ell}{m}=0,\,\forall
\ell>L_f$. For azimuthally symmetric function, where
$h(\unit{x}) = h(\theta, \phi) = h(\theta)$, we note that only zero-order spherical
harmonic coefficients of $h(\unit{x})$ are non-zero, that is,
$\shc{h}{\ell}{m} = 0$ for all $m\neq 0$.

For notational compactness, we express the spherical
harmonic $Y_{\ell}^m$ as $Y_n$ and spherical harmonic coefficient $
\shc{f}{\ell}{m}$ as $\shc{f}{n}{}$. That is, as a function of a
single integer index $n$ instead of two integer indices ${\ell}$ and
$m$, using the \emph{one-to-one} mapping
\begin{equation}
	\label{Eq:mapping}
	({\ell},m)\leftrightarrow n, \quad n = {\ell}^2+{\ell}+m, \quad \ell
		= \lfloor \sqrt{n}\rfloor, \quad m = n - \lfloor \sqrt{n}
		\rfloor(\lfloor \sqrt{n}\rfloor-1),
\end{equation}
where $\lfloor\,\cdot\,\rfloor$ denotes the integer floor
function. Using this mapping, the \emph{spectral response} can be
denoted by $\bv{f} =
\parn[\big]{\shc{f}{0}{}, \, \shc{f}{1}{}, \,  \shc{f}{2}{}, \,
\dotsc}'$. If $f(\unit{x})$ is band-limited in degree to $L_f$, then $\bv{f} =
\parn[\big]{\shc{f}{0}{},\shc{f}{1}{}, \shc{f}{2}{}, \dotsc,
\shc{f}{N_f}{}}'$, where $N_f = L_f^2+2L_f$.

\subsection{Rotations on the Sphere}

Here, we only define the rotation for azimuthally symmetric signals
on the 2-sphere. We refer the reader elsewhere~\cite{kennedy:book} for
comprehensive details for rotations defined on the 2-sphere. Define
the rotation operator $\rotopz$ with $\unit{y} = \unit{y}(\vartheta,\varphi)$, which rotates the azimuthally
symmetric functions $h(\unit{x})$ by $\vartheta\in [0,\,\pi]$ about
$y-$axis followed by a $\varphi\in[0,\,2\pi)$ rotation about
$z-$axis. Under this rotation operation $\rotopz$, the spherical
harmonic coefficients of the rotated signal are related to those of
the original signal through~\cite{kennedy:book}
\begin{equation}
	\label{eqn:rotation_symm}
	\parn[\big]{\rotopz h}_\ell^m = \sqrt{\frac{4\pi}{2\ell+1}}
		\conj{Y_\ell^m(\vartheta,\varphi)} \shc{h}{\ell}{0}, \quad \unit{y} = \unit{y}(\vartheta,\varphi).
\end{equation}

\subsection{Convolution}

There are different definitions of spherical convolution available
in the literature~\cite{Yeo:2008,Kennedy:2011,Sadeghi:2012}. In this
paper, we follow the definition of convolution presented as harmonic
multiplication~\cite{Kennedy:2011}
\begin{equation}
\label{Eq:conv_def}
    \parn[\big]{h\circledast f}(\unit{y}) \dfn  \sum_{{\ell}=0}^{\infty} \sum_{m=-\ell}^\ell \shc{h}{\ell}{m} \shc{f}{\ell}{m} Y_\ell^m(\unit{y}),
\end{equation}
where $\shc{h}{\ell}{m}=\innerp[\big]{h}{Y_{{\ell}}^m}$ and
$\shc{f}{\ell}{m}=\innerp[\big]{f}{Y_{{\ell}}^m}$. We also note that
\begin{equation}
\label{Eq:conv_def_azimuthally}
    \shc{h\circledast f}{\ell}{m}  = \innerp[\big]{h\circledast
    f}{Y_{{\ell}}^m} = \shc{h}{\ell}{m} \shc{f}{\ell}{m}.
\end{equation}

\subsection{Stochastic Processes on the Sphere}
Let $z(\unit{x})$ denotes a realization of a zero-mean, Gaussian,
\emph{anisotropic}, random process on the 2-sphere with spectral
domain representation
\begin{equation}
	z(\unit{x}) = \sum_{n}^{\infty} \shc{z}{n}{} Y_n(\unit{x}).
\end{equation}
Since we assume the process to be zero-mean and Gaussian, we take
$\shc{z}{n}{}$ to be complex-valued jointly normal random variables
and therefore the covariance between the spherical harmonic
coefficients completely characterizes the process.  Define
$\mathbf{C}_Z$ as the spectral covariance matrix with entries of the
form $\mathbf{C}_Z^{nr}$ given by
\begin{equation}
\label{Eq:process_prop_spectral}
	\mathbf{C}_Z^{nr} \dfn \expect[\big]{\shc{z}{n}{}\overline{\shc{z}{r}{}}}
\end{equation}
as a measure of the covariance between spectral coefficients
$\shc{z}{n}{}$ and $\shc{z}{r}{}$ of the realizations of the processes,
where $\expect{\cdot}$ denotes the expectation operator.  Also,
define the spatial covariance function $C_Z(\unit{x}, \unit{y}) =
\expect{z(\unit{x}) \overline{z(\unit{y})}}$ which
relates the process values at two spatial positions.  Given the spectral
covariance matrix $\mathbf{C}_Z$, the spatial covariance
$C_Z(\unit{x}, \unit{y})$ can be determined as
\begin{equation}
\label{Eq:process_prop_spatial}
	C_Z(\unit{x}, \unit{y}) = \sum_{n=0}^{\infty} \sum_{r=0}^\infty
	\mathbf{C}_Z^{nr}Y_n(\unit{x})  \overline{Y_r(\unit{y})}
\end{equation}

We note that the covariance matrix $\mathbf{C}_Z$, or the covariance
function $C_Z(\unit{x}, \unit{y})$, given in
\eqref{Eq:process_prop_spectral}, or
\eqref{Eq:process_prop_spatial}, is the most general
characterization of the random Gaussian (anisotropic) process on the
sphere~\cite{Hitczenko:2012}, which can be used to describe some special
processes on the 2-sphere. For example, spectral covariance matrices
for isotropic process and azimuthally symmetric process are given by
\begin{align*}
	\text{Isotropic process:} &\quad \mathbf{C}_Z^{nr} = C_\ell
		\delta_{nr},\quad (\ell,m)\leftrightarrow n, \\
	\text{Azimuthally-symmetric process:} &\quad \mathbf{C}_Z^{nr} =
		C^m_{\ell p} \delta_{mq}, \quad (\ell,m)\leftrightarrow
		n,\,(p,q)\leftrightarrow r,
\end{align*}
for some real $C_{\ell}$ and $C^m_{\ell p}$ which can be used in
conjunction with \eqref{Eq:process_prop_spatial} to determine
spatial covariance functions as
\begin{align*}
	\text{Isotropic process:}&\quad C_Z(\unit{x}, \unit{y}) =
		\sum_{\ell=0}^{\infty }\frac{2\ell+1}{4\pi}C_\ell
		P_\ell^0(\unit{x}\cdot\unit{y}), \\
	\text{Azimuthally-symmetric process:}&\quad C_Z(\unit{x},
		\unit{y}) = \sum_{m={-\infty}}^{\infty}\, \sum_{\ell=|m|}^{\infty}
		\, \sum_{p=|m|}^{\infty} C^m_{\ell p}\, N_\ell^m N_p^m \,
		P_\ell^m(\cos\theta) P_p^m(\cos\theta).
\end{align*}

\section{Spatio-Spectral Filtering Framework}
\label{sec:ss_filtering}

In this paper, we are interested in the filtering and modification
of signals in the joint spatio-spectral domain. We use the spatially
localized spherical harmonic transform~(SLSHT) distribution as
a representation of signal in the spatio-spectral
domain~\cite{Khalid:2012,Khalid2:2012,Khalid:2013}. The SLSHT
distribution is obtained using spatially localized spherical
harmonic transform~(SLSHT) for signals on the 2-sphere, which is analogous to the
short-time Fourier transform~(STFT). We
consider the filtering framework~\cite{Khalid2:2012} shown in
\figref{fig:block_concept}, where the SLSHT distribution of the
input signal is first obtained, then the SLSHT distribution is
processed in the joint spatio-spectral domain to yield the filtered
distribution and transformed back to the spatial domain using the SLSHT
inverse operation.

\subsection*{Spatially Localized Spherical Harmonics
Transform~(SLSHT) Distribution}\label{Sec:model_SLSHT}

Analogous to the STFT, the SLSHT can be
defined as a set of windowed spherical harmonic
transforms~\cite{Simons:1997,Khalid:2012} to represent the signal
jointly in the spatio-spectral domain. Mathematically, the SLSHT of
the signal $f(\unit{x})$ using an azimuthally symmetric window function
$h(\unit{x})$, evaluated at point $\unit{y} =
\unit{y}(\vartheta,\varphi)$, degree $\ell$ and order $m$, is
defined as
\begin{equation}
\label{Eq:stft_spatial}
    g_f(\unit{y};{\ell},m) \equiv g_f(\unit{y};n) \dfn
    \intsph
    \parn[\big]{\rotopz h}(\unit{x})f(\unit{x}) \conj{Y_{\ell}^m(\unit{x})}
    \,ds(\unit{x}), \quad
    (\ell,m)\leftrightarrow n.
\end{equation}
which forms the \emph{SLSHT distribution} given by
\begin{equation}
\label{Eq:stft_matrix_def}
    \bv{g}_f(\unit{y})\dfn \parn[\big]{g_f(\unit{y};0),\, g_f(\unit{y};1),\, g_f(\unit{y};2),\,
        \dotsc}',
\end{equation}
as a representation of signal in the spatio-spectral domain. Unlike the
spherical harmonic coefficient $\shc{f}{\ell}{m}$, which is only a
function of degree $\ell$ and order $m$, the SLSHT provides a
spatially-varying spherical harmonic representation of the signal, that is,
$g_f(\unit{y};{\ell},m)$ is a function of the spatial localization
$\unit{y}$, degree $\ell$ and order $m$.

\begin{remark}
For mathematical simplification, we assume that the azimuthally
symmetric window function $h(\unit{x})$ is band-limited with maximum
spectral degree $L_h$ and the spectral response is $\bv{h} =
\parn[\big]{\shc{h}{0}{},\,\shc{h}{1}{},\,\shc{h}{2}{},\dotsc,
\shc{h}{N_h}{}}'$, where $N_h = L_h^2+2L_h$. We further assume that
the window function is unit energy normalized, that is,
$\innerp[\big]{h}{h}=1$.
\end{remark}


Summarizing the mathematical formulations developed
elsewhere~\cite{Khalid:2012,Khalid2:2012}, we express the SLSHT
distribution component ${g}_f(\unit{y};n)$ as
\begin{align}
\label{Eq:stft_spectral}
    {g}_f(\unit{y};n)
    	&= \sum_{r=0}^{N_h} \parn[\Big]{\sum_{c=0}^{\infty} \shc{f}{c}{}
			\sqrt{\frac{4\pi}{2p+1}} \shc{h}{p}{0} \,T(c;r;n)}
			\conj{Y_p^q(\unit{y})} \nonumber \\
		&= \sum_{c=0}^{\infty} \shc{f}{c}{}\, \psi_{n,c}(\unit{y}),
\end{align}
with
\begin{equation}
\label{Eq:kernel_element}
    \psi_{n,c}(\unit{y})
    	= \sum_{r=0}^{N_h} \sqrt{\frac{4\pi}{2p+1}}\conj{Y_p^q(\unit{y})}
			\shc{h}{p}{0}\,T(c;r;n),  \quad (p,q)\leftrightarrow r
\end{equation}
where
\begin{equation}
\label{Eq:triple_prod_abc}
	T(c;r;n) = T(r;c;n) \dfn \intsph Y_r(\unit{x}) Y_c(\unit{x})
		\conj{Y_n(\unit{x})} \,ds(\unit{x})
\end{equation}
denotes the spherical harmonic triple product and its explicit
expression is provided in \appref{App:triple}.

\begin{remark} 
\label{rem:bandlim}
Comparing \eqref{Eq:stft_spectral} with
\eqref{Eq:f_expansion}, we note that each distribution component
${g}_f(\unit{y};n)$ is band-limited with maximum spectral degree
$L_h$~(band-limit of the window function).
\end{remark}

\subsection*{Spatially Varying Filtering}

The SLSHT distribution represents the spatially-varying spectral
representation of a signal as a function of both spatial location
$\unit{y}$ and degree and order $\ell,m$, and therefore offers the
opportunity to filter the signal in the joint
spatio-spectral domain. For this purpose, we define set of
functions as the filter function
\begin{equation*}
	\bv{\zeta}(\unit{y}) \dfn
	\parn[\big]{\zeta(\unit{y};0),\,\zeta(\unit{y};1),\,\zeta(\unit{y};2),\,\dotsc}'
\end{equation*}
in the spatio-spectral domain, with each element
$\zeta(\unit{x};n)$ for $0\leq n$ be a finite-norm, square
integrable band-limited function on the 2-sphere with the maximum
spectral degree $L_{\zeta_n}$ and spherical harmonic expansion
$\zeta(\unit{y};n) = \sum_{r=0}^{N_{\zeta_n}}
\shc{\zeta_n}{r}{}Y_r(\unit{y})$, where $N_{\zeta_n} =L_{\zeta_n}^2
+ 2L_{\zeta_n}$.

Define the filtered distribution
\begin{equation*}
    \bv{v}(\unit{y})\dfn \parn[\big]{v(\unit{y};0),\,v(\unit{y};1),\,v(\unit{y};2),\,\dotsc}',
\end{equation*}
where each filtered distribution
component ${v}(\unit{y};n)$ is obtained as a convolution of the SLSHT
component distribution $g_f(\unit{y};n)$ and the corresponding element $\zeta(\unit{y};n)$ of the filter function
$\bv{\zeta}(\unit{y}) $, that is,

\begin{equation}\label{Eq:conv_spatio_spectral}
{v}(\unit{y};n) = \big(g_f(\cdot\,;n) \circledast
		\zeta(\cdot\,;n)\big)(\unit{y}).
\end{equation}

\noindent By defining
\begin{equation*}
	G_f(r;n) = \innerp[\big]{g_f(\cdot\,;n)}{Y_r}
\end{equation*}
and using the formulation of convolution given in \eqref{Eq:conv_def}, we express the filtered distribution component ${v}(\unit{y};n)$ in \eqref{Eq:conv_spatio_spectral} as
\begin{equation}
\label{Eq:filtered_dist_component}
	{v}(\unit{y};n) = \smashoperator{\sum_{r=0}^{\min (N_h,N_{\zeta_n})}} G_f(r;n) \shc{\zeta_n}{r}{}Y_r(\unit{y}).
\end{equation}

\begin{remark} 
Following Remark \ref{rem:bandlim} and noting the fact that each component
${v}(\unit{y};n) $ of the filtered distribution is obtained as a result
of convolution of $g_f(\unit{y};n)$ and $\zeta(\unit{y};n)$, which
corresponds to the multiplication in the spectral domain as
described in \eqref{Eq:conv_def_azimuthally}, the summation in
\eqref{Eq:filtered_dist_component} is truncated at the minimum of $N_h$ and
$N_{\zeta_n}$. For simplicity, we assume that each filter
component $\zeta(\unit{y};n)$ is band-limited with maximum spectral
degree $L_h$, that is, $L_{\zeta_n}=L_h$ for $n\in\{1,2,\dotsc\}$.
\end{remark}

\subsection*{Inverse SLSHT}

Here we present inverse SLSHT to obtain the signal $\tilde
s(\unit{x}) \in \lsph$ having spectral response $\bv{s}$ and the SLSHT distribution $\bv{g}_{\tilde
s}(\unit{y}) $, which approximates $\bv{v}(\unit{y})$ in the least
squares sense.  We can determine such a signal $\tilde s(\unit{x})$ in spectral domain as~\cite{Khalid2:2012}
\begin{equation}\label{Eq:thm_state}
    \shc{\tilde
s}{c}{} = \sum_{n=0}^{\infty} \intsph
\overline{\psi_{c,n}}(\unit{y}) \, {v}(\unit{y};n)\,ds(\unit{y}).
\end{equation}

\section{Problem Formulation}
\label{sec:problem}

With all of the necessary mathematical background presented, we
formulate the problem under consideration, namely the estimation of
signals on the 2-sphere from noisy observations. We consider the
complex valued signals on the 2-sphere, contaminated by anisotropic
additive noise for which the statistical properties are only known.
Assume that the signal of interest is $s(\unit{x})$, contaminated by
additive noise $z(\unit{x})$. Given
\begin{equation}\label{Eq:input_signal_sum}
f(\unit{x}) = s(\unit{x}) + z(\unit{x}),
\end{equation}
we consider the problem to find the \emph{optimal} (in a sense to be defined) estimate of the
signal $s(\unit{x})$.

The noise is considered to be a realization of zero mean, Gaussian, anisotropic,
random process on the 2-sphere with known covariance
matrix $\mathbf{C}_Z$. We also assume that the signal of interest
$s(\unit{x})$, itself, may be a realization of zero mean, Gaussian,
anisotropic, random process for which the covariance
matrix $\mathbf{C}_S$ is also known. Furthermore, we assume that
noise and signal are uncorrelated, that is, $\expect{ s(\unit{x})
\overline{z(\unit{x}})}=0$~(or equivalently $\expect{ \shc{s}{n}{}
\overline{\shc{z}{r}{}}}=0$).
%
We note that the problem under consideration, which is to estimate the signal
$s(\unit{x})$ from $f(\unit{x})$, is equivalent to finding the best
estimate $\shc{\tilde{s}}{n}{}$ of the spherical harmonic expansion
coefficients $\shc{s}{n}{}$ forming the spectral response, $\bv{s}$ of the signal $s(\unit{x})$.

We also need to clarify what we understand by ``optimal"
in this context; we seek an estimate
$\tilde{s}(\unit{x})$~(or $\bv{\tilde{s}}$), which minimizes the
mean-square error~(MSE) between the estimate and the desired signal.
Define the MSE $\mathcal{E}_s$, that is, the expected energy of the estimation
error as
\begin{align}
\label{Eq:problem_form_MSE}
	\mathcal{E}_s &= \expect[\Big]{ \norm[\big]{s(\unit{x}) - \tilde{s}(\unit{x})}^2 }
		\nonumber \\
		&= \expect[\big]{ (\bv{s} - \bv{\tilde{s}})^H (\bv{s} - \bv{\tilde{s}})},
\end{align}
where $(\cdot)^H$ denotes the Hermitian and the second equality which quantifies the error between the estimate
and the signal in the spectral domain, follows from the orthonormality of
spherical harmonics.

For isotropic processes, there exist simple spectral domain
formulations for optimal filters which minimizes the mean-square
error between the estimate and the signal~\cite{Wei:2011} or
equivalently the variance of the unbiased estimate~\cite{McEwen:2008}. In our present formulation the signal and noise processes can be \emph{anisotropic}, and this puts us in the realm of designing \emph{spatially-varying} optimal filters for the estimation of a signal from its noise contaminated version. We design such an optimal filter in the
next section.

%


\begin{figure}[tbp]
    \centering
    \includegraphics[scale=0.85]{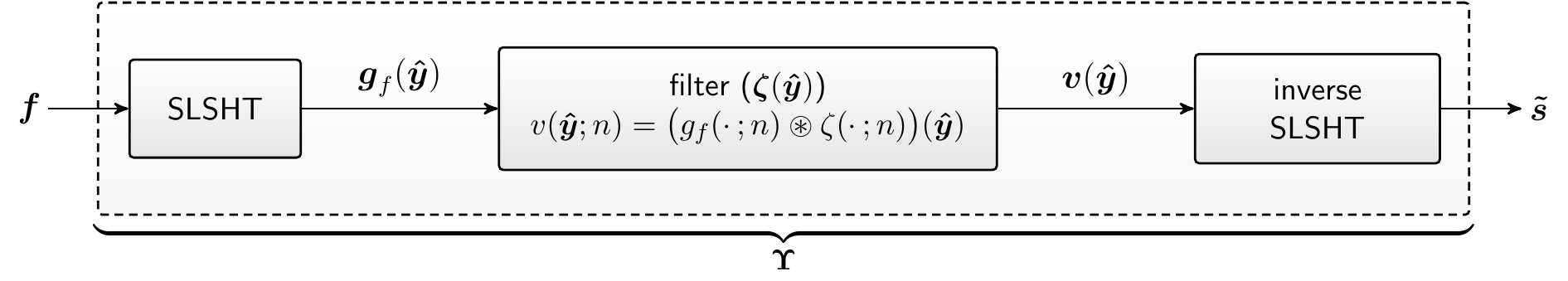}\vspace{2mm}
    \caption{The spatio-spectral filtering framework:
    $\bv{f}$ is the spectral response of the signal $f(\unit{x})$ on the 2-sphere, $\bv{g}_f(\unit{y})$ is corresponding SLSHT distribution
    in the spatio-spectral domain, which convolves
    with the filter function $\bv{\zeta}(\unit{y})$ to generate filtered distribution $\bv{v}(\unit{y})$ and $\bv{\tilde s}$ is the spectral response of the
    estimated signal $\tilde s(\unit{x})$, obtained using inverse SLSHT.  The transformation between the spectral responses is linear and given by the spatio-spectral transformation matrix $\bv{\Upsilon}$.}
    \label{fig:block_concept_ss}
\end{figure}

\section{Optimal Filtering in Spatio-Spectral Domain}
\label{sec:optimal_filtering}

We consider the spatio-spectral filtering framework presented in \secref{sec:ss_filtering} to estimate the
signal $s(\unit{x})$ from $f(\unit{x})$.  As depicted in \figref{fig:block_concept_ss}, let $f(\unit{x})$, given in \eqref{Eq:input_signal_sum},
be the input signal to the filtering framework, where the
filter function $\bv{\zeta}(\unit{y})$ convolves with the SLSHT
distribution $\bv{g}_f(\unit{y})$ of the input signal $f(\unit{x})$, generating a filtered distribution
$\bv{v}(\unit{y})$ for which we obtain the signal estimate $\tilde
s(\unit{x})$ using inverse SLSHT. In this setting, the estimation problem formulated in the previous section reduces to
determine such an optimal filter $\bv{\zeta}(\unit{y})$ that minimizes
the mean square error, given in \eqref{Eq:problem_form_MSE}, between the signal
estimate ${\tilde s(\unit{x})}$ and the uncontaminated signal signal
$s(\unit{x})$. This is equivalent to seeking an optimal filter which minimizes the
MSE between the filtered distribution
$\bv{v}(\unit{x})$ and the SLSHT distribution $\bv{g}_s(\unit{y})$
of the uncontaminated signal $s(\unit{x})$. Define the spatio-spectral MSE $\mathcal{E}_{ss}$ as
\begin{align}\label{Eq:mse_dist}
	{\mathcal{E}_{ss}}
	&= \expect[\big]{\sum_{n=0}^{\infty} \norm[\big]{
		{v}(\unit{y};n)  - {g}_s(\unit{y};n) }^2 } \nonumber \\
	&= \expect[\big]{ \sum_{n=0}^{\infty} \norm[\big]{ \big(g_f(\cdot\,;n) \circledast
		\zeta(\cdot\,;n)\big)(\unit{y}) - {g}_s(\unit{y};n) }^2 }.
\end{align}
Now we derive an optimal filter $\bv{\zeta}(\unit{y})$  which minimizes MSE ${\mathcal{E}_{ss}}$ given in \eqref{Eq:mse_dist} and present the result in the form
of following theorem.

\begin{theorem}[Optimal filter in the spatio-spectral domain]
\label{thm:optfilt}
Let $f(\unit{x}) = s(\unit{x}) + z(\unit{x})$ be the sum of a desired signal $s(\unit{x})$
and a noise $z(\unit{x})$. Further, let $\mathbf{C}_S$ and $\mathbf{C}_Z$ be the covariance matrices
as defined in \eqref{Eq:process_prop_spectral} for signal and noise processes, respectively.
For $f(\unit{x})$ as an input to the spatio-spectral filtering framework presented in \secref{sec:ss_filtering}~(shown in
\figref{fig:block_concept_ss}) with filter function $\bv{\zeta}(\unit{y}) =
	\parn[\big]{\zeta(\unit{y};0),\,\zeta(\unit{y};1),\,\zeta(\unit{y};2),\,\dotsc}'$, the spectral response, $\shc{\zeta_n}{r}{} = \innerp[\big]{\zeta(\cdot\,;n)}{Y_r}$ with $0\leq r \leq (N_h = L_h^2+2L_h)$, of the optimal filter which minimizes the mean-square error ${\mathcal{E}_{ss}}$, formulated in
\eqref{Eq:mse_dist}, between the filtered distribution $\bv{v}(\unit{y})$ and the
SLSHT distribution $\bv{g}_s(\unit{y})$, is given by
\begin{equation}
\label{Eq:thm:statement}
\shc{\zeta_n}{r}{} =  \frac{
\sum\limits_{c=0}^{\infty}\sum\limits_{c'=0}^{\infty}H(c;r;n)
H(c';r;n) \mathbf{C}_S^{cc'}
}{\sum\limits_{c=0}^{\infty}\sum\limits_{c'=0}^{\infty}H(c;r;n)
H(c';r;n)( \mathbf{C}_S^{cc'} + \mathbf{C}_Z^{cc'})}
\end{equation}
for $n$ and $r$ such that
$\sum\limits_{c=0}^{\infty}\sum\limits_{c'=0}^{\infty}H(c;r;n)
H(c';r;n)
( \mathbf{C}_S^{cc'} + \mathbf{C}_Z^{cc'} )>0$ and is zero otherwise. Here $H(c;r;n)$ is given by
\begin{equation*}
	H(c;r;n) =
		(-1)^q \sqrt{\frac{4\pi}{2p+1}} \shc{h}{p}{0}\,T(c;p,-q;n), \quad (p,q)\leftrightarrow r.
\end{equation*}
\end{theorem}

\noindent{\bf Proof:}
We first define
\begin{equation*}
G_s(r;n) = \innerp[\big]{g_s(\cdot\,;n)}{Y_r}, \quad
V(r;n) = \innerp[\big]{v(\cdot\,;n)}{Y_r},
\end{equation*}
where $0 \leq r \leq (N_h = L^2+2L_h)$, following Remark \ref{rem:bandlim} and
\begin{equation}\label{Eq:dfn_H}
H(c;r;n) = \innerp[\big]{\psi_{n,c}(\cdot\,)}{Y_r},
\end{equation}
following \eqref{Eq:kernel_element}. We can
express $G_f(r;n)$ and $G_s(r;n)$ using the formulation of
${g_f(\unit{y};n)}$ given in \eqref{Eq:stft_spectral} and \eqref{Eq:dfn_H} as
\begin{align}
G_f(r;n) &= \sum_{c=0}^{\infty} \shc{f}{c}{}H(c;r;n) =
\sum_{c=0}^{\infty} \left(\shc{s}{c}{} +
         \shc{z}{c}{}\right)H(c;r;n),
\nonumber \\
G_s(r;n) &= \sum_{c=0}^{\infty} \shc{f}{c}{}H(c;r;n) =
\sum_{c=0}^{\infty} \shc{s}{c}{} H(c;r;n),
\nonumber
\end{align}
and noting the effect of convolution in the spectral domain given in
\eqref{Eq:conv_def_azimuthally}, $V(r;n)$ can be written as
\begin{align}
V(r;n) &=  G_f(r;n) \shc{\zeta_n}{r}{}.
\end{align}

\noindent Using these formulations, we write the MSE ${\mathcal{E}_{ss}}$ in
\eqref{Eq:mse_dist} as
\begin{align}\label{Eq:mse_dist_detailed}
{\mathcal{E}_{ss}} &= \expect[\bigg]{ \sum_{n=0}^{\infty}
	\norm[\big]{\sum_{r=0}^{N_h }\left( V(r;n) - G_s(r;n)\right) Y_r(\unit{y})}^2 }
	\nonumber \\
	&= \sum_{n=0}^{\infty} \sum_{r=0}^{N_h} \expect[\Big]{ \abs[\big]{ V(r;n) - G_s(r;n)}^2}
	\nonumber \\
	&= \sum_{n=0}^{\infty} \sum_{r=0}^{N_h} \sum_{c=0}^{\infty}\sum_{c'=0}^{\infty}H(c;r;n) H(c';r;n)
	\nonumber \\
	& \qquad \times \expect[\bigg]{
		\parn[\Big]{
			\shc{\zeta_n}{r}{} \parn[\big]{\shc{s}{c}{} + \shc{z}{c}{}} - \shc{s}{c}{} }
		\parn[\Big]{
			\overline{\shc{\zeta_n}{r}{}}  \parn[\big]{ \overline{\shc{s}{c'}{}} +
         \overline{\shc{z}{c'}{}}} - \overline{\shc{s}{c'}{}}  }}.
\end{align}
Here $\abs[\big]{(\cdot)}^2 \dfn (\cdot) \overline{(\cdot)}$.
Now, noting the fact that the signal and noise are uncorrelated,
that is $\expect{ \shc{s}{c}{} \overline{\shc{z}{c'}{}}}=0$ for all $c,
\,c'$, using
$\expect{\shc{s}{c}{}\overline{\shc{s}{c'}{}}}=\mathbf{C}_S^{cc'}$ and
$\expect{\shc{z}{c}{}\overline{\shc{z}{c'}{}}}=\mathbf{C}_Z^{cc'}$ and
setting the derivative of MSE ${\mathcal{E}_{ss}}$ in
\eqref{Eq:mse_dist_detailed}, with respect to
$\overline{\shc{\zeta_n}{r}{}}$, equal to zero,  we obtain the
result stated in \eqref{Eq:thm:statement} for spectral response of
the optimal filter $\bv{\zeta}(\unit{y})$.

\hfill\qed
%

\begin{remark} 
For the case when signal distortions are less than acceptable, we might require an optimal filter which minimizes a weighted MSE. The weighted MSE is obtained by
assigning weight $\alpha$ to the factor in the MSE given in \eqref{Eq:mse_dist_detailed} that is due to
the noise, and assigning weight $(1-\alpha)$ to the remaining factor. This amounts to replacing $\mathbf{C}_Z$ by $\alpha \mathbf{C}_Z$ and $\mathbf{C}_S$ by $(1-\alpha)\mathbf{C}_S$ in the solution of the optimal filter in \eqref{Eq:thm:statement}.
\end{remark}
%

The optimal filter signal estimate $\tilde
s(\unit{x})$ can be recovered form the spectral domain signal $\bv{\tilde s}$
using the inverse SLSHT operation. Using
\eqref{Eq:thm_state}, we obtain the estimate $\shc{\tilde s}{c}{}$
as
\begin{align}
\label{Eq:sig_inverse}
	\shc{\tilde s}{c}{}
		&= \sum_{n=0}^{\infty} \intsph \overline{\psi_{c,n}}(\unit{y})\,  {v}(\unit{y};n)\, ds(\unit{y})
	\nonumber \\
		&= \sum_{n=0}^{\infty} \sum_{c'=0}^{\infty} \sum_{r=0}^{N_h} \shc{f}{c'}{} \shc{\zeta_n}{r}{}\,H(c;r;n)\, H(c';r;n).
\end{align}
By defining the matrix $\mathbf{\Upsilon}$ with entries
\begin{align}
\mathbf{\Upsilon}^{c,c'}=  \sum_{n=0}^{\infty} \sum_{r=0}^{N_h} \shc{\zeta_n}{r}{}\,H(c;r;n)\, H(c';r;n).
\end{align}
the overall process of finding the desired signal estimate $\tilde s(\unit{x})$ with spectral response $\bv{\tilde s}$,
from the signal $f(\unit{x})$  with spectral response $\bv{f}$ can be expressed
as following linear transformation as shown in \figref{fig:block_concept_ss}, that is,
\begin{equation}
\label{Eq:sig_lin_tx}
	\bv{\tilde s} = \mathbf{\Upsilon}\bv{f}.
\end{equation}

The optimal filter in the spatio-spectral domain, derived in Theorem \ref{thm:optfilt},
extends the spectral domain formulation of optimal isotropic symmetric
filter~\cite{Klees:2008}, which is valid for isotropic processes, to the
broader class of anisotropic processes. Since we have defined
$G_s(r;n) = \innerp[\big]{g_s(\cdot\,;n)}{Y_r}$, we note
that the spatio-spectral domain can be parameterized by $n$ and
$r$~(recalling $n,\,r \in \mathbb{Z}^{+})$. With this
characterization of spatio-spectral domain at hand, the formulation
of optimal filter in the spatio-spectral domain given in
\eqref{Eq:thm:statement} has a simple and intuitively appealing
interpretation. For the values of $n$ and $r$: 1) where only the
contribution of signal is present, that is, $G_z(r;n)\approx0$, the
optimal filter is approximately one, thus allows the signal to pass
through without distortion; 2) where only the noise contribution is
present, that is, $G_s(r;n)\approx0$, or contribution of both signal
and noise are not present, that is,  $G_s(r;n)\approx
G_z(r;n)\approx0$, the optimal filter is approximately zero and
therefore it suppresses the noise; and 3) where the contribution of both
noise and signal are present, the optimal filter performs a
weighting that depends on the signal and noise contribution in the
spatio-spectral domain at the respective values of $n$ and $r$.

\begin{remark} 
Since the SLSHT distribution as a spatio-spectral representation is
obtained using a window function, the overall optimal filtering is
dependent on the chosen window function, the choice of which entails
a trade-off between its resolution in the spatial and spectral
domains. In time-frequency analysis, the analogous problem is resolved
by extending the use of multiple windows~(a set of orthogonal windows),
the concept originally introduced by Thompson~\cite{Thomson:1982},
in obtaining the STFT representation for the filtering of
non-stationary signal in the time-frequency
domain~\cite{Kozek:1996}. The same multi-window concept is
used for the estimation of power spectrum for signals on the 2-sphere
from the observations available over some spatially limited
region~\cite{Wieczorek:2005}. We anticipate that the similar concept
can be employed to improve the performance of the proposed optimal
filter. However, further exploration in this direction is a subject of future work.
\end{remark}

\begin{figure*}[t]
    \centering
    \subfloat[Desired signal $s(\unit{x})$]{
        \includegraphics[width=0.45\textwidth]{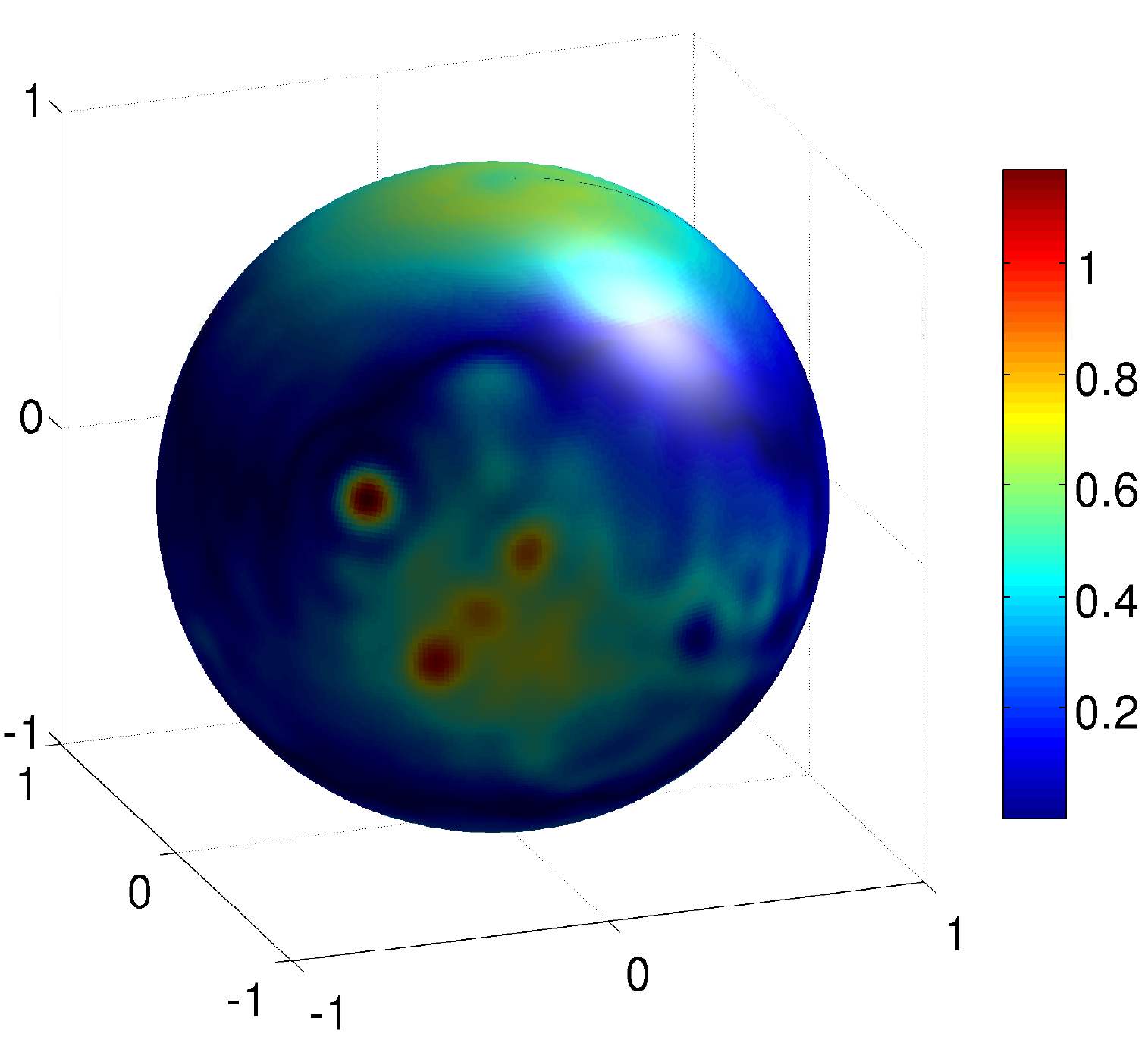}}
        \hspace{7mm}
    \subfloat[Noise contaminated signal $f(\unit{x}) = s(\unit{x})+z(\unit{x})$]{
        \includegraphics[width=0.45\textwidth]{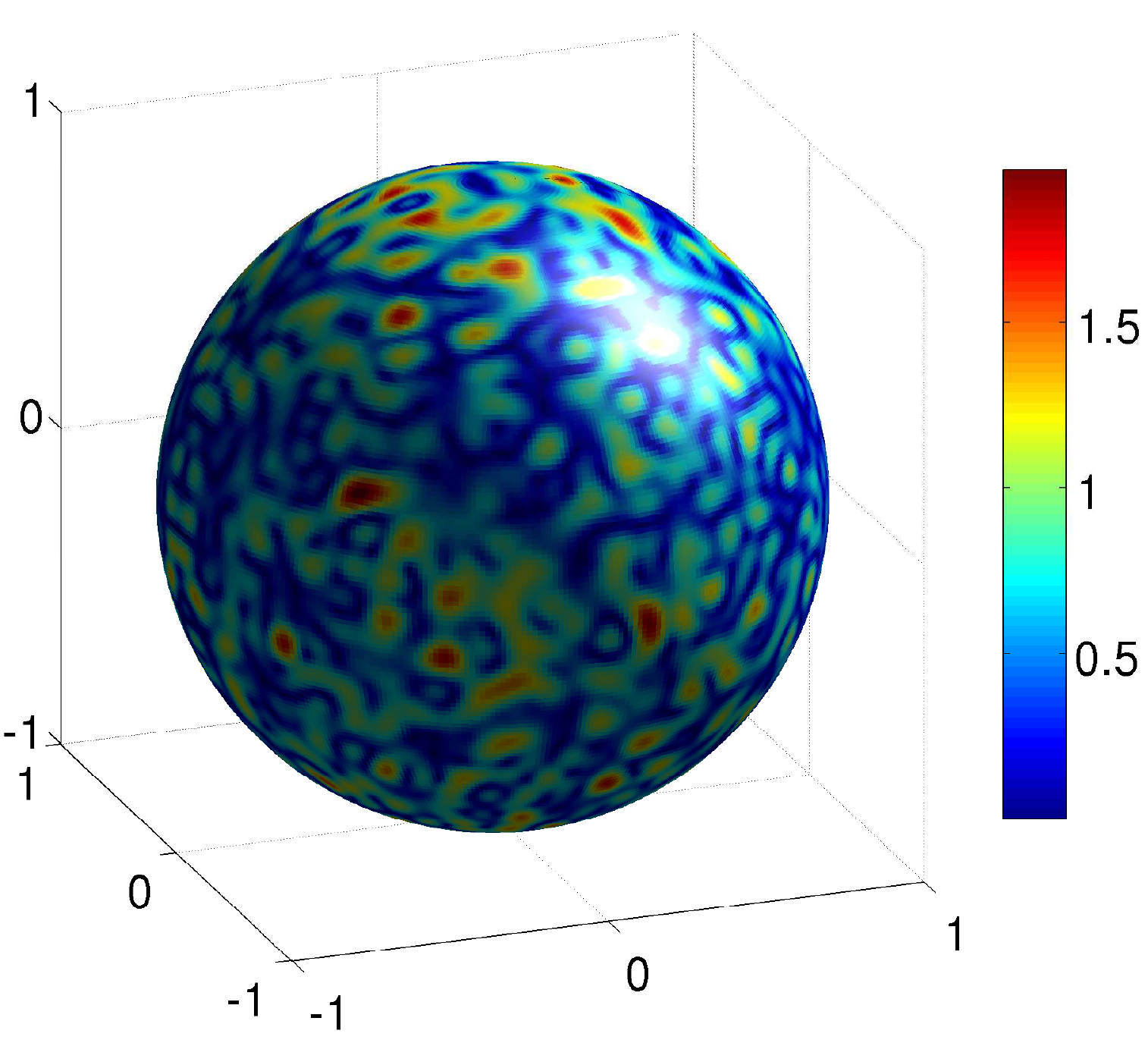}} \\[3mm]

    \caption{(a) The magnitude of the desired signal~(Mars topographic map) $s(\unit{x})$, band-limited with maximum spectral degree $L_s = 36$ and
    (b) the magnitude of the noise contaminated signal  $f(\unit{x}) =
s(\unit{x}) + z(\unit{x})$ with $\snr^f = -0.04$dB. }
\label{fig:input_signals}
\end{figure*}

\begin{figure*}[t]
    \centering
    \subfloat[${\abs[\big]{G_s(r;n)}}$]{
        \includegraphics[width=0.48\textwidth]{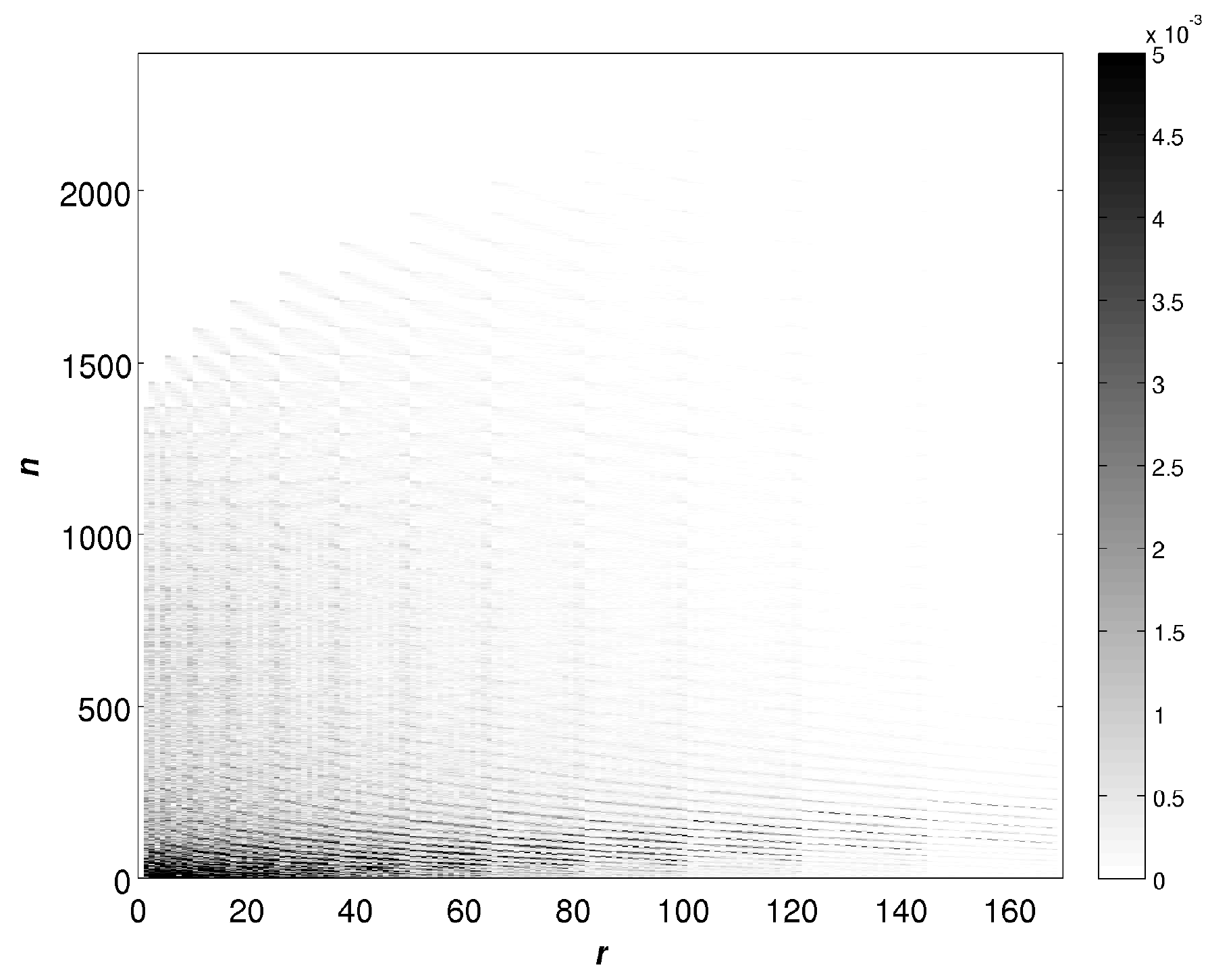}} \hspace{3mm}
    \subfloat[${\abs[\big]{G_f(r;n)}}$]{
        \includegraphics[width=0.48\textwidth]{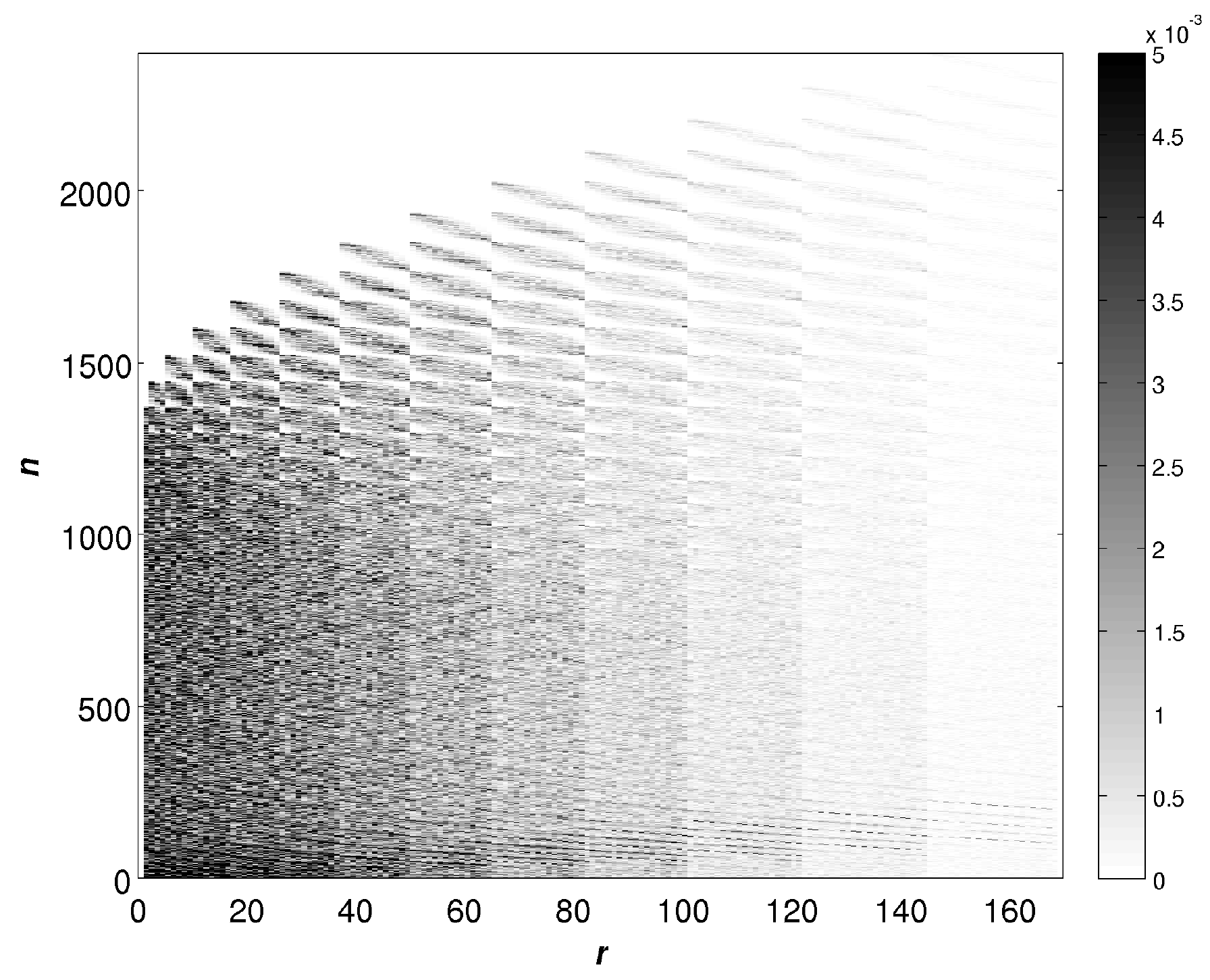}} \\
    \subfloat[${\abs[\big]{\shc{\zeta_n}{r}{}}}$]{
       \includegraphics[width=0.48\textwidth]{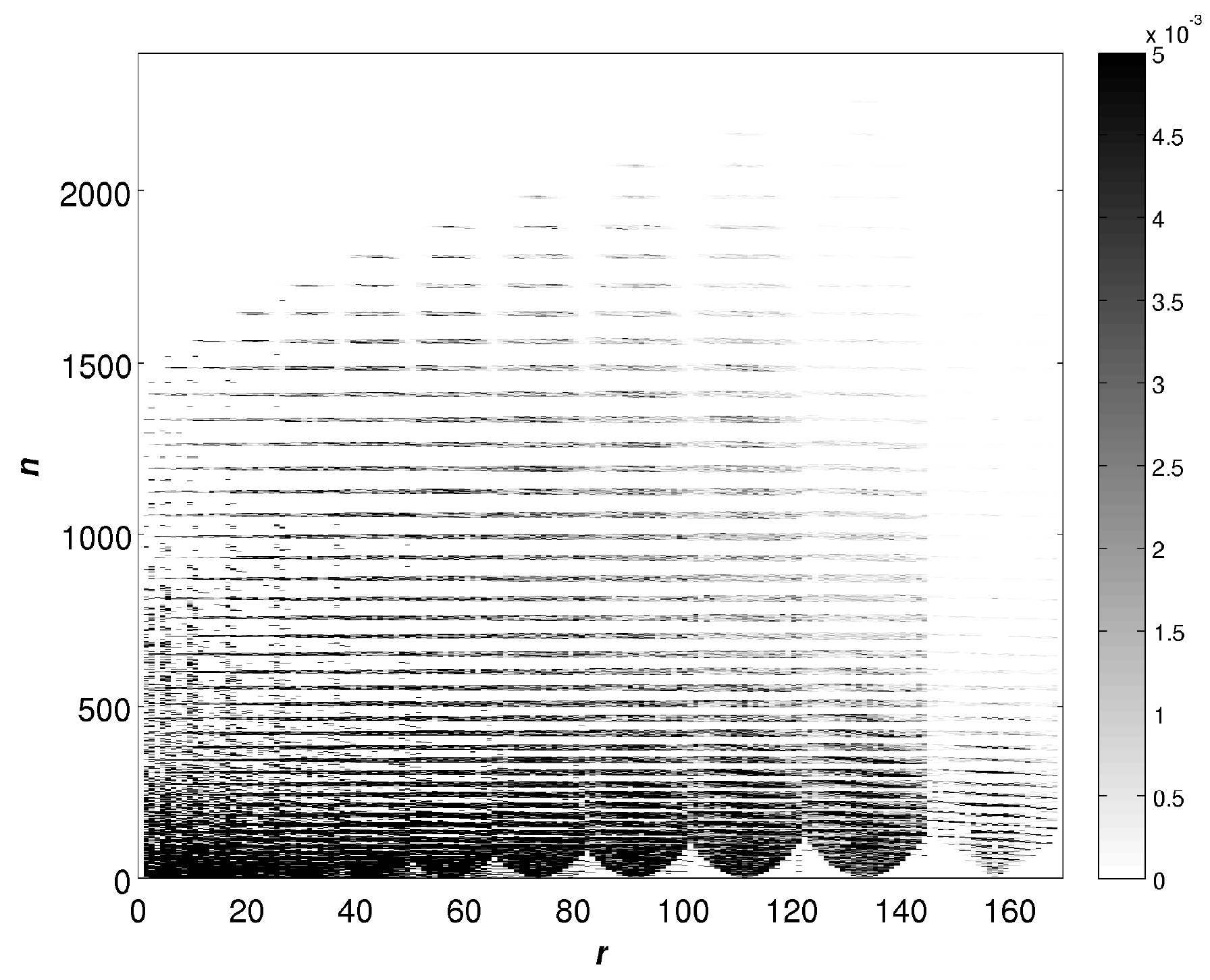}}\hspace{3mm}
    \subfloat[${\abs[\big]{V(r;n)}}$]{
       \includegraphics[width=0.48\textwidth]{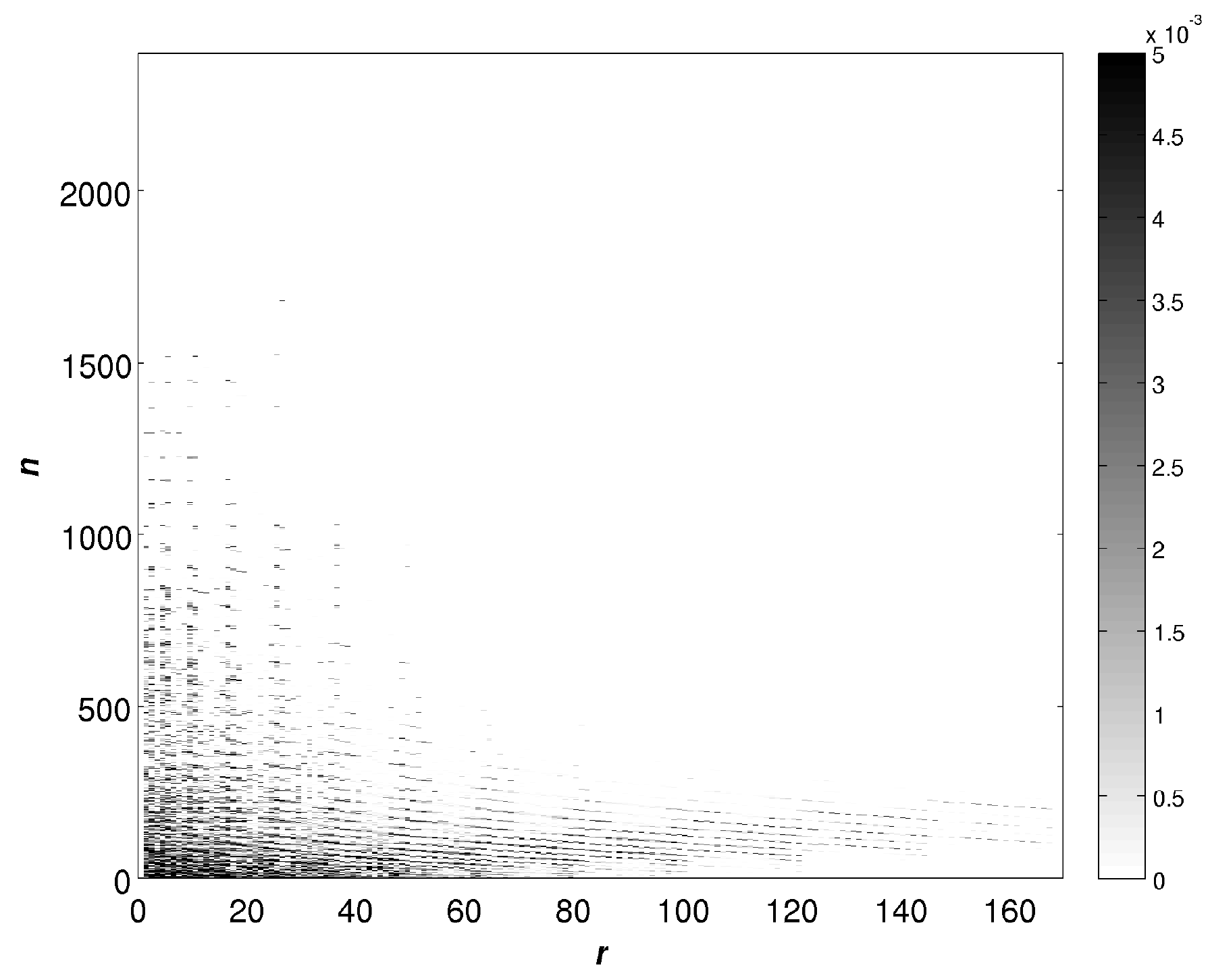}} \\[3mm]
	\caption{The SLSHT distributions $\bv{g}_s(\unit{y})$~(for the desired signal) and $\bv{g}_f(\unit{y})$~(for the noise contaminated signal) are shown in (a) $\abs{G_s(r;n)}$ and (b) $\abs{G_f(r;n)}$. The optimal filter $\bv{\zeta}(\unit{y})$ derived in the spectral domain using the result in Theorem \ref{thm:optfilt} is shown as (c) $\abs{\shc{\zeta_n}{r}{}}$ and (d) the filtered distribution $\bv{v}(\unit{y}) $ is shown as $\abs{V(r;n)}$.  All of the plots are shown for $0 \leq r \leq N_h$ and $0 \leq n \leq N_g$.}
\label{fig:distributions}
\end{figure*}

\begin{figure}
  \centering
  \includegraphics[width=0.45\textwidth]{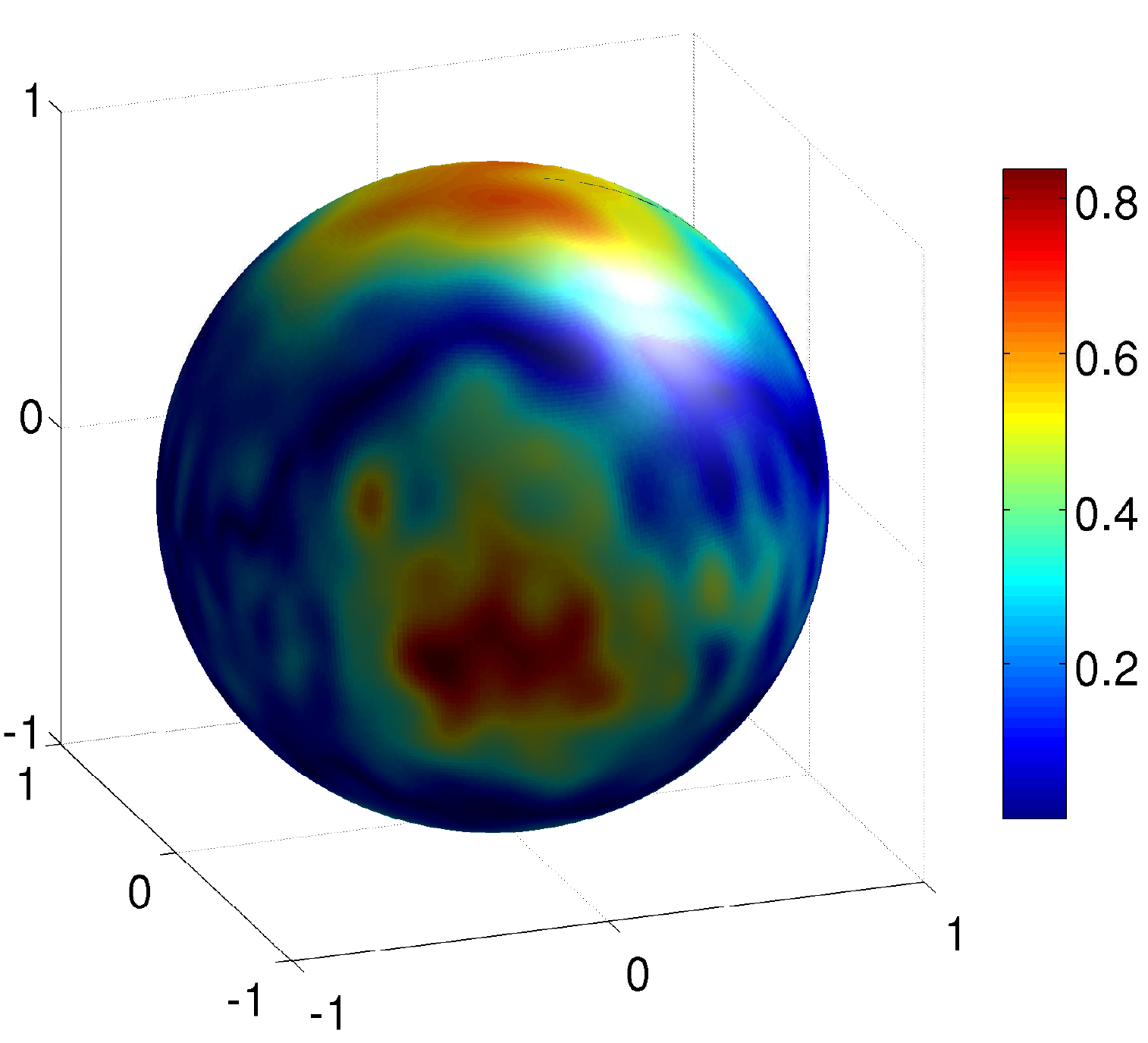}
  \caption{The magnitude of the estimated signal $\tilde s(\unit{x})$, obtained from filtering the signal $f(\unit{x})$ shown in \figref{fig:input_signals}(b) using the
  proposed optimal filter in the spatio-spectral domain. The SNR of the output signal is $\snr^{\tilde s} = 24.7$dB. }
  \label{fig:recovered_signal}
\end{figure}

\section{Optimal Filter Design Example}
\label{sec:example}

We apply the proposed optimal filter in the spatio-spectral domain
to enhance a signal that has been corrupted by anisotropic noise. The objective
here is to demonstrate the performance of the proposed filter. We
use signal-to-noise ratio~(SNR) as a measure to quantify the
performance of the optimal filter. If $s(\unit{x})$ denotes
the desired signal, we define the SNR for the signal $f(\unit{x})$
as
\begin{equation*}
	\snr^{f} = 20 \log \frac{\norm[\big]{ s(\unit{x}) }}{\norm[\big]{ f(\unit{x}) - s(\unit{x}) }}.
\end{equation*}
in order to quantify the performance of the proposed optimal filter
in the spatio-spectral domain. The implementation of optimal filtering framework is implemented
in \texttt{MATLAB}, using the \texttt{MATLAB} interface of the
\texttt{SSHT}\footnote{\url{http://www.jasonmcewen.org/}} package. The SLSHT distribution is
computed using the efficient method proposed in literature\cite{Khalid:2013}.

We consider the Mars topographic map~(height above geoid) as a
desired signal $s(\unit{x})$, which is synthesized using the
spherical harmonic model of the topography of
Mars.\footnote{\url{http://www.ipgp.fr/~wieczor/SH/}} For
convenience, the desired signal is made band-limited with band-limit
$L_s=36$. We normalize the signal such that it has unit energy, that
is, $\norm{s}=1$. The desired signal $s(\unit{x})$ is shown in
\figref{fig:input_signals}(a). The noise is assumed to be a realization of zero mean,
Gaussian, \emph{anisotropic}, random process on the 2-sphere with
known covariance matrix $\mathbf{C}_Z$. The covariance matrix is
constructed as $\mathbf{C}_Z = \mathbf{A}\,\mathbf{A}^H$, where the
entries of matrix $\mathbf{A}$ are complex with real and imaginary
parts uniformly distributed in the interval $[0,1]$. The covariance
matrix is then normalized such that the noise process has unit
energy within the band-limit of the desired signal $L_s$, that is,
$\sum_{n=0}^{L_s^2+2L_s}\mathbf{C}_Z^{rr} = 1$. The noise
$z(\unit{x})$ is then generated in the spectral domain using the
covariance matrix. It should be noted that the generated noise is
complex valued.

The sum of the desired signal and the noise, that is, the noise contaminated signal $f(\unit{x}) =
s(\unit{x}) + z(\unit{x})$, is shown in \figref{fig:input_signals}(b).  It is applied as an input to the spatio-spectral filtering framework shown in \figref{fig:block_concept_ss}. The SLSHT distribution of the input
signal is obtained using the azimuthally symmetric band-limited
eigenfunction window~\cite{Khalid:2012} with band-limit $L_h=12$ and
$99\%$ energy concentration in a polar cap region of central angle
$\pi/6$. The SLSHT distributions $\bv{g}_s(\unit{y})$~(desired) and
$\bv{g}_f(\unit{y})$~(input) are plotted as $G_s(r;n)$ and
$G_f(r;n)$ in \figref{fig:distributions}(a) and \figref{fig:distributions}(b), respectively. Since both the desired signal and the window function are band-limited signals on the 2-sphere,
both $G_s(r;n)$ and $G_f(r;n)$ are only defined for $0 \leq r \leq N_h$ and $0 \leq n \leq N_g$, where $N_g = (L_f+L_h)^2 + 2(L_f+L_h)$. The
optimal filter $\shc{\zeta_n}{r}{}$, defined in
Theorem \ref{thm:optfilt}, is shown in \figref{fig:distributions}(c) for different values of $n$
and $r$. The optimal filter is applied to the SLSHT distribution
$\bv{g}_f(\unit{y};n)$ to yield the filtered distribution
$\bv{v}(\unit{y})$ shown in \figref{fig:distributions}(d) as $V(r;n)$, which is then
inverted, using \eqref{Eq:sig_inverse}, to obtain the estimated
signal $\tilde s(\unit{x})$ shown in \figref{fig:recovered_signal}.

Since both the desired signal and the noise process
covariance matrix are normalized to contain unit energy within
the band-limit $L_s$, then the expected SNR is $0$dB.  For the realization in this
example
the actual input SNR is $\snr^f \approx
-0.04$dB and the output SNR is $\snr^{\tilde s} = 24.7$dB. We note that the
proposed optimal filter, taking into account the signal and noise
statistics, provides significant SNR improvement. We have provided
example merely to demonstrate the capability of the proposed
spatio-spectral optimal filtering framework. The more rigorous
analysis of the performance of the filter and the application to
real data such as Cosmological Microwave Background~(CMB)
maps~\cite{planck:bluebook} and Gravity Recovery and Climate
Experiment~(GRACE) gravity models are the subjects of future work.

\section{Conclusions}
\label{sec:conclusions}

We have developed an optimal filter for the estimation of
signals on the 2-sphere corrupted by the noise, for the case when both
the signal and noise are realizations of anisotropic processes on the
sphere.  The optimal filter is based on transformation in the spatio-spectral domain
recently presented in the literature, and minimizes the
mean-square error between the estimated signal and desired signal.
The proposed optimal filter, unlike filters
available in literature, takes into account the anisotropic
properties of random processes on the 2-sphere. Finally, we provided
an example to demonstrate the capability of proposed optimal filter.
For future work we highlight two open problems: employing an analogy of the multiple window
method~\cite{Thomson:1982} used in time frequency
analysis~\cite{Kozek:1996} for the improvement of performance of the
proposed optimal filter; and the application of the proposed optimal filter to
the real data.

\appendix

\section{Mathematical Background}

\subsection{Spherical Harmonics}
\label{App:maths}

The spherical harmonic function, $Y_{\ell}^m(\unit{x}) =
Y_{\ell}^m(\theta, \phi)$, for degree ${\ell} \geq 0$ and order $
|m| \leq {\ell}$ is defined as~\cite{Sakurai:1994}
\begin{equation*}
    Y_{\ell}^m(\theta, \phi) =
    N_{\ell}^{m}P_{\ell}^{m}(\cos\theta)\,e^{im\phi},
\end{equation*}
where $N_{\ell}^{m}$ is the normalization factor given by
\begin{equation*}
    N_{\ell}^{m} \dfn \sqrt{\frac{2{\ell}+1}{4\pi}\frac{({\ell}-m)!}{({\ell}+m)!}},
\end{equation*}
such that $\innerp[\big]{Y_{\ell}^m}{ Y_{p}^{q}}=\delta_{\ell p}
\delta_{mq}$, where $\delta_{mq}$ is the Kronecker delta function:
$\delta_{mq} = 1$ for $m=q$ and is zero otherwise. $P_{\ell}^{m}(x)$
is the associated Legendre function defined for degree $\ell$ and
order $0 \leq m \leq \ell$ as
\begin{align*}
    P_{\ell}^{m}(x)
        &= \frac{(-1)^m}{2^{\ell} {\ell}!} (1-x^2)^{m/2} \frac{d^{{\ell}+m}}{dx^{{\ell}+m}}
            (x^2-1)^{\ell} \\
    P_{\ell}^{-m}(x)
        &= {(-1)^m} \frac{({\ell}-m)!}{({\ell}+m)!} P_{\ell}^m(x),
\end{align*}
for $|x|\leq 1$.

\subsection{Spherical Harmonics Triple Product}
\label{App:triple}

Using the Wigner-$3j$ symbols~\cite{Sakurai:1994}, the spherical
harmonic triple product $T(u;r;n)$ can be written using the mappings
$(a,b)\leftrightarrow c$, $(p,q)\leftrightarrow r$ and
$(\ell,m)\leftrightarrow n$ as
    \begin{equation*}
        T(c;r;n) = (-1)^m \,
            \sqrt{\frac{(2a+1)(2p+1)(2\ell+1)}{4\pi}} \,
        \begin{pmatrix}
          a & p & \ell \\
          0 & 0 & 0
        \end{pmatrix}
        \begin{pmatrix}
          a & p & \ell \\
          b & q & -m
        \end{pmatrix}.
    \end{equation*}


\acknowledgments     

This work was supported under the Australian Research Council's Discovery Projects funding scheme (Project No.~DP1094350).

\bibliography{SLSHT_new} 
\bibliographystyle{spiebib}   

\end{document}